\documentstyle[12pt,epsfig]{article}
\def\ct#1{{\cal #1}}

\newcommand{\nn}{\nonumber}

\newcommand{\be}{\begin{equation}}
\newcommand{\ee}{\end{equation}}
\newcommand{\bea}{\begin{eqnarray}}
\newcommand{\eea}{\end{eqnarray}}

\begin{document}
\pagestyle{empty} 
\begin{flushright}
CPHT-S054,0500 \\
FTUV-IFIC-00-0512\\
LPT-ORSAY/00-48 \\
RM3-TH/00-5 \\
ROMA 1293/00 \\
\today
\end{flushright}
\centerline{\LARGE \bf $K^{0}$--$\bar K^{0}$ Mixing with Wilson 
Fermions}
\vskip 0.2cm
\centerline{\LARGE \bf  without  Subtractions}
\vskip 0.7cm
\centerline{\large {D.~Becirevic$^{a}$, Ph.~Boucaud$^{b}$, V. 
Gim\'enez$^c$, V.~Lubicz$^d$, }}
\centerline{\large {G. Martinelli$^{a,e}$, J.~Micheli$^b$,  
M.~Papinutto$^{b,f}$}}
\vskip 0.3cm
\centerline{\sl $^a$ Dip. di Fisica, Univ. di Roma ``La Sapienza'' and
INFN, Sezione di Roma,}
\centerline{\sl P.le A. Moro 2, I-00185 Roma, Italy.}
\smallskip
\centerline{\sl $^{b}$ Laboratoire de Physique Th\'eorique (LPT), Universit\'e de 
Paris-Sud,}
\centerline{\sl  B\^atiment 210, 91405 Orsay, France.}
\smallskip
\centerline{\sl $^c$ Dep. de Fisica Teorica and IFIC, Univ. de Valencia,}
\centerline{\sl Dr. Moliner 50, E-46100, Burjassot, Valencia, Spain.}
\smallskip
\centerline{\sl $^d$ Dip. di Fisica, Univ. di Roma Tre and INFN,}
\centerline{\sl Sezione di Roma, Via della Vasca Navale 84, I-00146 Rome, Italy.}
\smallskip
\centerline{\sl $^e$ Centre de Physique Th\'eorique de l'\'Ecole Polytechnique,}
\centerline{\sl 91128 Palaiseau Cedex, France.}
\smallskip
\centerline{\sl $^f$ Dip. di Fisica, Univ. di Pisa and INFN,}
\centerline{\sl Sezione di Pisa, Via Buonarroti 2, I-56100 Pisa, Italy.}
\vskip 0.4 cm
\date{}
\abstract{ By using suitable Ward identities,
we show that it  is possible to compute $K^{0}$--$\bar K^{0}$
mixing  without  
subtracting the  terms generated by  explicit chiral symmetry 
breaking present in  Wilson-like lattice actions. The accuracy 
in the determination of the amplitudes is of ${\cal O}(a)$, which is 
the best one attainable in the absence of improvement.}

\vskip 0.4cm
{\small PACS: 13.75Lb,\ 11.15.Ha,\ 12.38.Gc. \par}
\vskip 0.2 cm 
\newpage
\pagestyle{empty}\clearpage
\setcounter{page}{1}
\pagestyle{plain}

\newpage
\pagestyle{plain} \setcounter{page}{1}

\section{Introduction}
\label{sec:introduction}
A key ingredient in the study of CP violation in the Standard Model 
is the theoretical prediction of the $K^0$--$\bar K^0$  
mixing amplitude. This involves the computation
of  the $\Delta S = 2$ matrix element 
\be \langle \bar K^0 \vert
O^{\Delta S = 2} \vert K^0 \rangle \equiv \frac{8}{3} f_{K}^{2} m_{K}^{2}
B_{K} \ee 
of the operator
\be
O^{\Delta S = 2}\equiv O_{1} = ({\bar s}^A \gamma_{\mu} (1-\gamma_5) 
d^A)({\bar s}^B 
\gamma_{\mu}   (1-\gamma_5)
d^B) \, , 
\label{eq:ods2}
\ee
where $s$ and $d$ stand for strange and down quarks and $A,B$ are 
colour indices.

In general, important information on the physics beyond the Standard Model, 
such as various SUSY extensions (MSSM, NMSSM,...),
can be obtained  by studying  $\Delta F = 2$
transitions (see \cite{Bsusy1} and references therein for a discussion).
For neutral kaons, such processes require, besides $O_{1}$,
 also the knowledge of the matrix elements of the operators
(we adopt here the notation of ref.~\cite{Bsusy1})
\bea 
O_{2} = ({\bar s^A} (1-\gamma_5) d^A) ({\bar s^B} (1-\gamma_5) d^B) \nn \\
O_{3} =({\bar s^A} (1-\gamma_5) d^B) ({\bar s^B} (1-\gamma_5) d^A) \nn \\
O_{4} =({\bar s^A} (1-\gamma_5) d^A) ({\bar s^B} (1+\gamma_5) d^B)
\label{eq:bsusy} \\
O_{5} =({\bar s^A} (1-\gamma_5) d^B) ({\bar s^B} (1+\gamma_5) d^A) \nn
\eea

\begin{figure}
\vspace*{-3.9cm}
\begin{center}
\begin{tabular}{@{\hspace{-0.7cm}}c c c}
&\epsfxsize9.0cm\epsffile{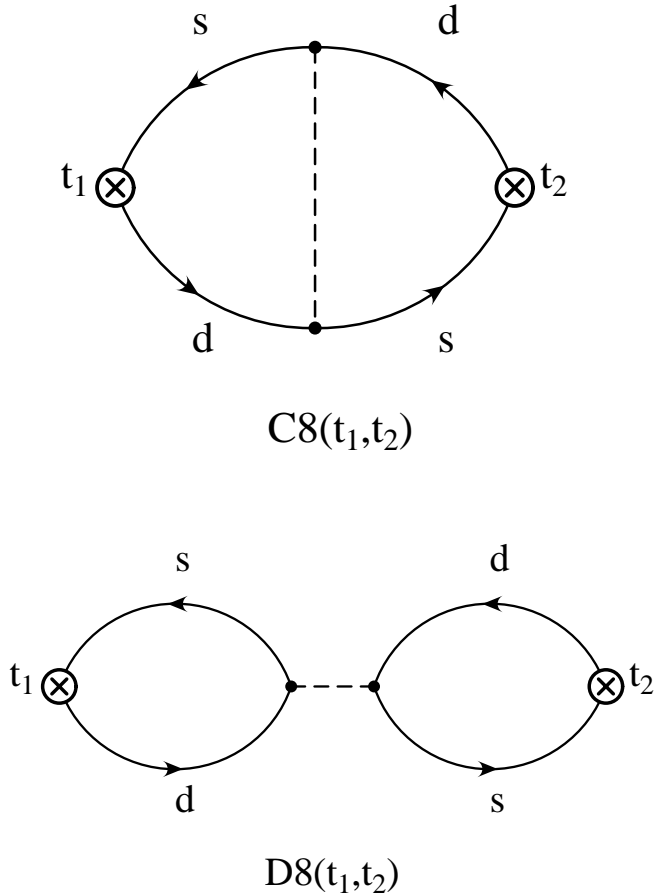} &  \\
\end{tabular}
\caption{\label{fig:eight}{\sl Connected and disconnected eight diagrams.
The dashed line represents the four-fermion operator.}}
\end{center}
\vspace*{-.3cm}
\end{figure}

On the lattice,  the matrix elements of the four-fermion operators above  are 
typically extracted from 
the large-time asymptotic behaviour of three-point correlation functions of the form $\langle 
K^{0}_P(x_{1}) O^{\Delta S = 2}(0) K^{0}_P(x_{2}) \rangle$,  where $K^{0}_{P}€$  are pseudoscalar sources
with suitable quark flavour, $K^{0}_{P}(x)=\bar d^A(x) \gamma_{5} s^A(x)$. 
Expressed in terms of traces of quark propagators,  these correlation functions correspond to
the so-called ``eight"- shaped quark diagrams given in  fig.~\ref{fig:eight}.

In order to obtain the physical amplitudes, it is necessary to 
compute the matrix elements of the renormalized operators 
corresponding to those defined in eqs.~(\ref {eq:ods2}) and 
(\ref{eq:bsusy}).  With Wilson-like fermions, the renormalization 
procedure is complicated by the presence of explicit chiral-symmetry
breaking ($\chi SB$) in the lattice fermion action: because of the Wilson term,
dimension-six operators belonging to different chiral representations can mix with 
each other. 
\par 
The testing ground for the restoration of chiral symmetry has been the chiral behaviour of
$\langle \bar K^0 \vert O^{\Delta S = 2} \vert K^0 \rangle$, which,
if properly renormalized, vanishes when the $K$-meson becomes 
massless~\cite{cabigel}. Although several  attempts with
Wilson fermions~\cite{GAVELA}--\cite{lello} have given
reasonable measurements of $B_K$, it remains true that the control
of the renormalization of the relevant operator is rather
problematic~\footnote{$B_K$ has also been obtained with staggered
fermions~\cite{SHARPE}--\cite{kil} mainly in the quenched approximation (see
ref.~\cite{Sharpe98} for a review). The (surviving) chiral symmetry
in the staggered fermion formalism ensures the vanishing
of the relevant matrix element in the chiral limit.}.

The root of the problem is the operator subtraction outlined above.
$O^{\Delta S =2}$ mixes with other operators $O_i$ of the same
dimension but with ``wrong na\"{\i}ve chirality". Thus, the ($\mu$-dependent)
$K^0$--$\bar K^0$ matrix element of the
renormalized operator $\hat O^{\Delta S = 2}$ is given in
terms of the ($a$-dependent) bare matrix elements by:
\be
\langle \bar K^0 \vert \hat O^{\Delta S =2} (\mu) \vert K^0 \rangle = 
\lim_{a \to 0} \langle \bar K^0 \vert Z^{\Delta S=2}_0 (a\mu,g_0^2) \left[ 
O^{\Delta S =2}(a) + \sum_i \Delta_i (g_0^2) O_i (a) \right]
\vert K^0 \rangle
\label{eq:mix}
\ee
The overall renormalization constant $Z^{\Delta S=2}_0 (a\mu,g^2_0)$ 
is logarithmically divergent and its determination does not affect the 
chiral behaviour of the matrix element.  The constants $\Delta_i(g^2_0)$'s are 
finite mixing  coefficients which depend on the lattice bare coupling 
(also expressed as $\beta = 6/g_0^2$) only~\cite{bochicchio}. They have been  either computed
in perturbation 
theory  (PT)~\cite{MARTIWdraper}--\cite{FREZZOTTIBORRELLI}, 
or fixed  by using the non-perturbative (NP) method  of 
refs.~\cite{NP}--\cite{bibbia},
or determined  using  the relevant Ward identities (WI) on quark states  (QWI)
\cite{jlqcd_lat96}~\footnote{
Another method,  
suggested in refs.~\cite{DI=1/2,marti98}, is the use of 
gauge-invariant WI on hadronic states.  This method has never  been implemented 
though.}.

When PT is used  for the calculation of the mixing coefficients $\Delta_i$,
the mass dependence of the renormalized-operator matrix 
element, $\langle \bar K^0 \vert {\hat O}^{\Delta S = 2} \vert K^0 \rangle$,
shows large deviations from the expected chiral behaviour.
Past experience suggests, instead,  that  the chiral behaviour of
 $\langle \bar K^0 \vert {\hat O}^{\Delta S = 2} \vert K^0 \rangle$
is satisfactory if a non-perturbative method, 
either the NP renormalization~\cite{B_K}--\cite{altri} 
or the QWI~\cite{jlqcd_lat96}, is used. 

In spite of this progress, the determination of the mixing 
coefficients is a long and  painful procedure.  The 
non-perturbative renormalization techniques proposed in the past 
have their   limitations  and, for this reason, 
  an accurate and systematic study of
the uncertainties  in the determination of the mixing 
coefficients would be necessary before keeping the final error on the 
matrix element fully under control. 

In this paper, {\it we propose a new method which allows the  
calculation  of   $\Delta S=2$  amplitudes  with Wilson fermions without 
determining the  mixing coefficients. The method is based on the 
lattice Ward 
identities  and it is non-perturbative. It can be 
applied to any Wilson-like formulation of the action (Wilson action, 
tree-level improved,  NPI and alike). Its accuracy,
in the absence of improvement  (of the action and of the operators), is of
${\cal O}(a)$, which is the best possible accuracy,
attainable only  with a perfect 
determination of the mixing coefficients $\Delta_{i}$ in 
previous approaches.}

 The same   method can be used to compute   $\Delta I =3/2$ $K$-$\pi$  matrix elements, 
  e.g. those relative to the electro-penguin 
operators, and   $\Delta I =3/2$   parity-conserving  amplitudes in hyperon decays.  It 
fails, unfortunately, for  $\Delta I=1/2$ transitions because of 
the mixing with lower-dimension operators. 

An alternative approach, based on the twisted mass QCD formalism~\cite{grassi},
also allows the determination of the $\Delta S =2$ mixing  amplitude
without lattice subtractions~\cite{sint}.

\section{Description of the Method} 

\label{sec:method}
In this section, we show how it is possible to compute the physical  
$K^{0}$--$\bar K^{0}$ mixing amplitude without knowing, or determining,  
the mixing coefficients induced by the explicit $\chi SB$ of the lattice action. We first recall some basic notions 
about operator renormalization and mixing with Wilson fermions and 
then explain our proposal.
\subsection{Operator Renormalization with Wilson Fermions}
\label{subsec:renormalization}
Following \cite{bibbia} (see also \cite{GUPTA} and \cite{BERNARD2}), we
classify  the complete basis of dimension-six, four-fermion 
operators which mix under renormalization,
relying on general symmetry arguments based on the
vector-flavour symmetry, which survives on the lattice.
To this purpose it is convenient to consider separately
the parity-even and parity-odd parts of the operators in eqs.~(\ref{eq:ods2}) and (\ref{eq:bsusy}). 
Thus, for example, the  parity-even part of the operator $O_{1}=Q_1-\ct{Q}_1$
 is given by
\be Q_{1}=  O_{[VV+AA]} = ({\bar s}^A \gamma_{\mu} 
d^A)({\bar s}^B  \gamma_{\mu}   d^B)  +({\bar s}^A \gamma_{\mu} \gamma_5 
d^A)({\bar s}^B  \gamma_{\mu}   \gamma_5  d^B) \, , \label{eq:q1} \ee
whereas the parity odd is defined as 
\be \ct{Q}_{1}=  O_{[VA+AV]} = 2 ({\bar s}^A \gamma_{\mu} 
d^A)({\bar s}^B  \gamma_{\mu} \gamma_5   d^B)  \, . \label{eq:tq1} \ee
On the basis of ${CPS}$ symmetries,  it can been shown that
the renormalization of the  $\ct{Q}_{i}$ operators is  not 
affected by the explicit $\chi SB$ of the lattice 
action and proceeds exactly  as in the ``continuum'' theory. 
The  corresponding renormalization  matrix $\ct{Z}_{ij}$ (which 
obviously depends on the renormalization-scheme) is a
 block diagonal matrix~\cite{bibbia}:
\begin{equation}
\left(\begin{array}{c} 
\hat{\ct{Q}}_1 \\
\hat{\ct{Q}}_2 \\
\hat{\ct{Q}}_3 \\
\hat{\ct{Q}}_4 \\
\hat{\ct{Q}}_5
\end{array}\right)
=
\left(\begin{array}{rrrrr}
\ct{Z}_{11} & 0 & 0 & 0 & 0 \\
0 & \ct{Z}_{22} & \ct{Z}_{23} & 0 & 0 \\
0 & \ct{Z}_{32} & \ct{Z}_{33} & 0 & 0 \\
0 & 0 & 0 & \ct{Z}_{44} & \ct{Z}_{45} \\
0 & 0 & 0 & \ct{Z}_{54} & \ct{Z}_{55}
\end{array}\right)
\left(\begin{array}{c} 
\ct{Q}_1 \\
\ct{Q}_2 \\
\ct{Q}_3 \\
\ct{Q}_4 \\
\ct{Q}_5
\end{array}\right) \, .
\label{eq:renpv}
\end{equation}
Thus the lattice does not induce extra subtractions 
($\Delta_{i}$) for the 
parity-odd sector ($\ct{Q}_k$; $k=1,\dots,5$), since the mixing in eq.~(\ref{eq:renpv}) is the same as it 
would appear in the absence of $\chi SB$. In 
particular, $\ct{Q}_{1}$ renormalizes multiplicatively.

The parity-even sector, on the contrary, is not protected by  ${CPS}$ symmetries, and 
all the five relevant operators get mixed because of the 
$\chi SB$ of the lattice action.
In this case, it is convenient to separate the
operator mixing into two classes: i) the first 
which consists in correcting the operator mixing  induced by the breaking of
chiral symmetry; ii) the second is the
renormalization which survives in the continuum limit.
In the absence of explicit $\chi SB$, the  mixing structure is  the same
as the one  considered above for the parity-odd counterparts. 
The corresponding parity-even   operators ($\tilde Q_k$; $k=1,\dots,5$)
would renormalize according to:
\begin{equation}
\left(\begin{array}{c} 
\hat Q_1 \\
\hat Q_2 \\
\hat Q_3 \\
\hat Q_4 \\
\hat Q_5
\end{array}\right)
=
\left(\begin{array}{rrrrr}
Z_{11} & 0 & 0 & 0 & 0 \\
0 & Z_{22} & Z_{23} & 0 & 0 \\
0 & Z_{32} & Z_{33} & 0 & 0 \\
0 & 0 & 0 & Z_{44} & Z_{45} \\
0 & 0 & 0 & Z_{54} & Z_{55}
\end{array}\right)
\left(\begin{array}{c} 
\tilde Q_1 \\
\tilde Q_2 \\
\tilde Q_3 \\
\tilde Q_4 \\
\tilde Q_5
\end{array}\right)\, , 
\label{eq:renor_subt}
\end{equation}
where the $\tilde Q_i$ represent the bare  operators, 
which transform as elements of irreducible representations of the 
chiral group (obviously up to terms of ${\cal O}(a)$). 
In the presence of the Wilson term, the $\tilde Q_i$ are
 defined as 
\begin{equation}
\left(\begin{array}{c} 
\tilde Q_1 \\
\tilde Q_2 \\
\tilde Q_3 \\
\tilde Q_4 \\
\tilde Q_5
\end{array}\right)
=
\left(\begin{array}{c} 
Q_1 \\
Q_2 \\
Q_3 \\
Q_4 \\
Q_5
\end{array}\right)
+
\left(\begin{array}{rrrrr}
0 & \Delta_{12} & \Delta_{13} & \Delta_{14} & \Delta_{15} \\
\Delta_{21} & 0 & 0 & \Delta_{24} & \Delta_{25} \\
\Delta_{31} & 0 & 0 & \Delta_{34} & \Delta_{35} \\
\Delta_{41} & \Delta_{42} & \Delta_{43} & 0 & 0 \\
\Delta_{51} & \Delta_{52} & \Delta_{53} & 0 & 0
\end{array}\right)
\left(\begin{array}{c} 
Q_1 \\
Q_2 \\
Q_3 \\
Q_4 \\
Q_5
\end{array}\right) \, .
\label{eq:renpc_sub}
\end{equation}
In other words, first the lattice subtraction is performed, followed by the
 renormalization of the remaining logarithmic divergencies. The above mixing pattern is abbreviated, in matrix
form, as
\bea
\hat Q &=& Z \tilde Q \nn \\
\tilde Q &=& [I+\Delta]Q
\eea
where $I$ is the $5 \times 5$ unit matrix.
\subsection{$K^{0}$--$\bar K^{0}$ Mixing without Subtractions}
\label{subsec:mixing}
For the sake of illustration, we discuss here the determination of the $K^{0}$--$\bar 
K^{0}$ matrix element of the operator $O^{\Delta S=2}=O_{1}$ only. The 
extension to the other operators ($O_{i}$, $i={2,\dots 5}$) is 
straightforward.
 
Le us consider the Ward identities which can be derived from the 
$\tau_{3}$  axial rotation 
\bea  \delta u =  \gamma_{5} u \, , \quad  \delta \bar u = \bar u \gamma_{5} 
\, , \nonumber \\   \delta d = - \gamma_{5} d \, , \quad \delta  \bar 
d =  - \bar d \gamma_{5} \, , \label{eq:rot} \eea
where $u$ and $d$ are the up and down quarks taken with degenerate 
masses,  $m_{u}=m_{d}=m$. 
For further use, we  also introduce  the following bilinear operators~\footnote{ 
As kaon source, we  may as well 
use the fourth component of the  axial current $\bar d \gamma_{0} 
\gamma_{5} s$ instead of  the 
pseudoscalar density.} 
\bea \Pi^{0}(x) &=&  \bar d(x) \gamma_{5} 
d(x) -\bar u (x)\gamma_{5} u(x)  \, , \quad  K_{P}^{0}(t)= \sum_{\vec x} \bar d(\vec x,t) \gamma_{5} 
s(\vec x,t)  \, , \nonumber \\  K_{S}^{0}(t)&=& \sum_{\vec x} \bar d(\vec x,t)  
s(\vec x,t)\, ,\label{eq:sources} \eea
and the corresponding renormalized quantities $\hat K_P^{0}(t) =Z_{P}  
K_P^{0}(t)$ and  $\hat K_{S}^{0}(t) =Z_{S}  
K_{S}^{0}(t)$.
The useful Ward identity in our case is then~\cite{bochicchio}
\bea 
\langle \delta\left[ \hat K_{P}^{0}(t_{1}) \hat O_{1}(0)\hat  K_{P}^{0}(t_{2})\right]   
\rangle - \langle \delta\left[ S \right]\hat  K_{P}^{0}(t_{1}) \hat O_{1}(0) \hat K_{P}^{0}(t_{2}) 
\rangle = 0 \, , \label{eq:wi} \eea
where $\delta\left[ \dots \right]$ denotes the rotation of the 
argument of $\delta$ and $\delta \left[ S \right]$ is the rotation of the action 
under the transformation defined in eq.~(\ref{eq:rot})~\footnote{
In general, different choices of the flavour content of the operators,
sources and rotation  are possible or may be necessary. For example
for  the operators $O_4$ and $O_5$,  the suitable choice is 
$O_{4,5} = (\bar s \Gamma  u)(\bar s \Gamma d)$ ($\Gamma$ matrices are not
specified) with $K_1 = \bar d \gamma_5 s$ and $K_2 = \bar u \gamma_5 s)$
and the same rotation as in eq.~(\ref{eq:rot}).}.

\begin{figure}
\vspace*{-3.9cm}
\begin{center}
\begin{tabular}{@{\hspace{-0.7cm}}c c c}
&\epsfxsize7.0cm\epsffile{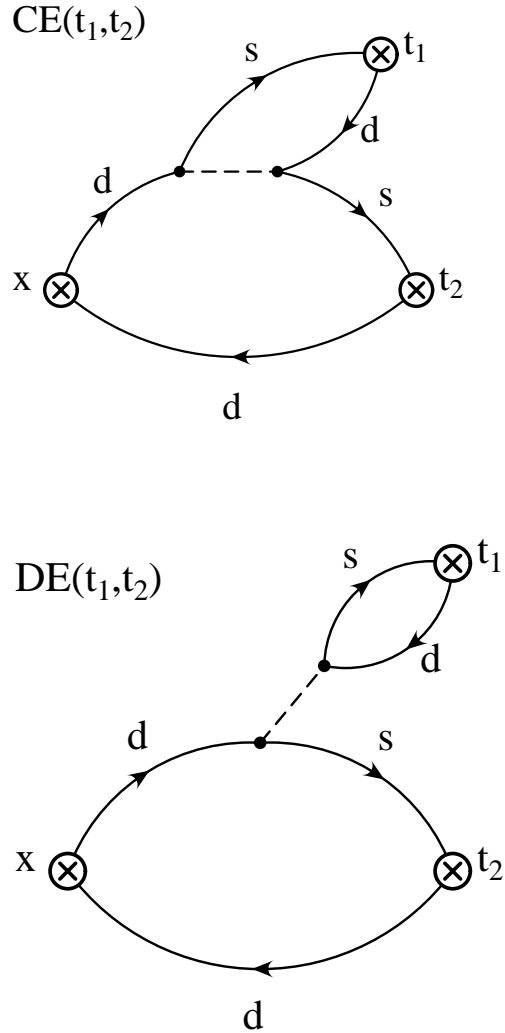} &  \\
\end{tabular}
\caption{\label{fig:dece}{\sl Connected and disconnected  emission diagrams.
The dashed line represents the four-fermion operator.
We denote as DE$(t_1,t_2)$ and $ C E (t_1,t_2)$
the diagrams where the $\bar K^0$ meson is emitted in 
$t_1$;  $DE(t_2,t_1)$ and $CE(t_2,t_1)$
the diagrams where the $\bar K^0$ meson is emitted in 
$t_2$.}}
\end{center}
\vspace*{-.3cm}
\end{figure}

In terms of the fields defined in eq.~(\ref{eq:sources}), and of the 
parity-even and parity-odd operators $Q_1$ and $\ct{Q}_1$, the Ward 
identity reads~\cite{bochicchio}
\bea &\,& 2 \langle \hat  K_{P}^{0}(t_{1}) \hat Q_{1}(0) \hat K_{P}^{0}(t_{2}) 
\rangle =  2 m \sum_{x} \langle \Pi^{0}(x) \hat  K_{P}^{0}(t_{1}) \hat 
\ct{Q}_{1}(0) \hat K_{P}^{0}(t_{2})  \rangle \nonumber \\
&\,& -   \langle  \hat  K_{S}^{0}(t_{1}) \hat 
\ct{Q}_{1}(0) \hat K_{P}^{0}(t_{2})  \rangle
-   \langle  \hat  K_{P}^{0}(t_{1}) \hat 
\ct{Q}_{1}(0) \hat K_{S}^{0}(t_{2})  \rangle   +{\cal O}(a) \, ,
 \label{eq:wi1} \eea
where, see eqs.~(\ref{eq:renpv})--(\ref{eq:renpc_sub}),  
\be \hat Q_{1} = Z_{11} \left( Q_{1} + \sum_{i=2,5} \Delta_{1i} 
Q_{i}\right)  \, ,  \quad \hat \ct{Q}_{1} = \ct{Z}_{11} \ct{Q}_{1} \, . \ee
The term on the l.h.s. of eq.~(\ref{eq:wi1}), corresponding to the 
rotation of the  operator $\ct{Q}_{1}$, is the quantity from which we  
may
extract the physical $K^{0}$--$\bar K^{0}$ mixing amplitude. The first term on the  r.h.s. 
corresponds to the insertion of the rotation of the action.  In terms of Feynman 
diagrams, it can be seen as the ``decay'' of a neutral (zero 
four-momentum) pion, 
$\pi^{0}$,  into two $\bar K^{0}$s under the action of $\ct{Q}_{1}$. 
 In the  $SU(2)$ isospin symmetric case ($m_{u}=m_{d}$), 
only the emission diagrams shown in 
fig.~\ref{fig:dece} must be considered. 
The last two terms in 
eq.~(\ref{eq:wi1}) correspond to the rotation of the pseudoscalar kaon 
sources. These terms are necessary to saturate  the Ward  identity.

One could  envisage the following method  to extract the coefficients 
$\Delta_{1i}$ and the ratio $\Delta_{11}=Z_{11}/\ct{Z}_{11}$, thus determining 
the subtracted operator $\tilde Q_{1}$. Since 
eq.~(\ref{eq:wi1}) must be satisfied for any value of $t_{1}$ and 
$t_{2}$, the Ward identity corresponds to a   system of linear 
  equations in the  unknown quantities $\Delta_{i}$ 
($i=11,12,\dots , 15$)~\cite{marti98}.  At least in principle, one has an 
independent equation for any assigned value of $t_{1}$ and 
$t_{2}$.  In practice, the equations  become dependent  when the Ward 
identity, for large values of $t_{1}$ and $t_{2}$, is saturated by  
the contribution of  the lowest lying states, namely the two $K$-mesons.

This method to extract the $\Delta_{i}$, which in practice may be
 very difficult to implement, is 
however unnecessary and a much easier procedure gives us directly the 
wanted quantity, namely  the correlation function of $\hat Q_{1}$. 
Let us consider the Ward identity (\ref{eq:wi1}) in the limit $t_{1} 
\to \infty$ and $t_{2} \to -\infty$ (in practice for large values of 
the time distances). In this limit, we may  safely neglect the last 
two terms of eq.~(\ref{eq:wi1}), because they correspond to the 
propagation of scalar states, which are exponentially suppressed with 
respect to the kaon contribution~\footnote{ This point can be explicitly
checked by computing $ \langle  \hat  K_{S}^{0}(t_{1}) \hat 
\ct{Q}_{1}(0) \hat K_{P}^{0}(t_{2})  \rangle$ in the same numerical 
simulation as the other correlation functions appearing in
eq.~(\ref{eq:wi1}).}. Then, up to exponentially suppressed 
terms,  we have ($\Delta_{11}= Z_{11}/\ct{Z}_{11}$ and we have divided all
the terms of the Ward identity by a factor of two)
\bea &\,& \lim_{t_{1} \to \infty,t_{2} \to -\infty}
\Delta_{11} \langle   K_{P}^{0}(t_{1}) \tilde  Q_{1}(0)  K_{P}^{0}(t_{2}) 
\rangle \to  \nonumber \\ &\,& 
\frac{\vert \langle 0 \vert  K_{P}^{0} \vert K^{0}\rangle \vert^{2}}{4 
m_{K}^{2}} e^{-m_{K}( t_{1} + \vert t_{2} \vert )} \times 
\ct{Z}^{-1}_{11} \langle  \bar K^{0} \vert \hat Q_{1} \vert K^{0} 
\rangle  \sim \nonumber \\ &\,&
  m \sum_{x} \langle \Pi^{0}(x)   K_{P}^{0}(t_{1})  
\ct{Q}_{1}(0)  K_{P}^{0}(t_{2})  \rangle  +{\cal O}(a) \,. \label{eq:final} \eea
Note that the last correlation function, $\langle \Pi^{0}(x)   K_{P}^{0}(t_{1})  
\ct{Q}_{1}(0)  K_{P}^{0}(t_{2})  \rangle$, is expressed  in terms 
of bare quantities only. Indeed a  single  constant  is sufficient to 
obtain the physical amplitude, namely   $\ct{Z}_{11}$ which relate the bare 
lattice parity-odd operator to the continuum one, renormalized in a 
specified renormalization scheme.  Obviously this constant cannot be determined 
from the Ward identity, since its values depend on the 
renormalization condition imposed to the renormalized 
operator~\cite{bochicchio,bibbia,jlqcd_lat96}.

\par In terms of the Feynman diagrams defined in figs.~\ref{fig:eight} and 
\ref{fig:dece}, eq.~(\ref{eq:final}) can be written as
\bea &\,& 2 \Bigl( C8(t_1,t_2) + D8(t_1,t_2) \Bigr) =  \\
&\,& 2 m \Bigl( CE(t_1,t_2)+CE(t_2,t_1) +DE(t_1,t_2) + DE(t_2,t_1) \Bigr)
+ {\cal O}(a) \, .
\nonumber \eea 
\par
In summary, the strategy to obtain the physical matrix element 
$\langle  \bar K^{0} \vert \hat Q_{1} \vert K^{0} \rangle$ is 
extremely simple:
\begin{itemize} 
\item one  computes the correlation function
\bea G_{3} (t_{1}, t_{2}) = m  \sum_{x} \langle \Pi^{0}(x)   K_{P}^{0}(t_{1})  
\ct{Q}_{1}(0)  K_{P}^{0}(t_{2})  \rangle \nonumber \eea  
of the bare parity-odd 
operator at large time distances $t_{1,2}$;  as  ``mass" $m$,
it is more convenient to use the quark mass 
defined as $m= Z_A \rho$, where
$\rho$ is defined using the axial Ward  identity~\cite{bochicchio,mm,luscher}
\bea 2 \rho = \frac{ \langle \partial_\mu  A_\mu \rangle }{ \langle P \rangle} 
\, . \nonumber  \eea
In the above equation 
$A_\mu$ and $P$ are the bare (eventually improved) axial current 
and pseudoscalar densities 
and the matrix elements are usually taken between the vacuum and a pion
(in our case the $\pi^0$)  at rest. 
$Z_A$ is the axial-current renormalization factor.
\item then one divides $G_{3}(t_{1}, t_{2})$ by the factor 
\bea {\cal F}(t_{1}, t_{2}) = \frac{Z_{5}}{4 
m_{K}^{2}} e^{-m_{K}( t_{1} + \vert t_{2} \vert )} \, , \nonumber \eea 
where $Z_{5}= \vert \langle 0 \vert  K_{P}^{0} \vert K^{0}\rangle 
\vert^{2} $, obtaining  the quantity 
${\cal R} =\ct{Z}^{-1}_{11} \langle   \bar K^{0} \vert \hat Q_{1} 
\vert  K^{0}  \rangle$. 
${\cal F}$ can  be readily computed from the study of the kaon two-point 
correlator $\langle K_{P}^{0}(t) K^{0 \dagger}_{P}(0) \rangle$.
Alternatively,   one  may construct the ratio
\bea R(t_{1}, t_{2}) = Z_{5} \frac{G_{3}(t_{1}, t_{2}) }{ 
\langle K_{P}^{0}(t_{1}) K^{0 
\dagger}_{P}(0) \rangle\langle K_{P}^{0}(t_{2}) K^{0 \dagger}_{P}(0) 
\rangle}   \to  {\cal R} \, , \nonumber \eea
at large time distances.
\item finally, the physical amplitude is given by 
\bea  \langle  \bar K^{0} \vert \hat Q_{1} \vert K^{0} 
\rangle = \ct{Z}_{11} \times  {\cal  R} \, .\nonumber \eea
The constant $\ct{Z}_{11}$ cannot be determined from the Ward 
identity and has to be fixed either using perturbation theory, or 
non-perturbatively on quark states~\cite{NP}--\cite{bibbia} or with the 
Schr\"odinger functional method~\cite{luscher}.
\end{itemize}
The application of the method discussed above to the  operators 
of eq.~(\ref{eq:bsusy}),  appearing in extensions of  the Standard 
Model, is so 
easy that it does not require further discussion.  The same approach 
can be used to compute the $\Delta I=3/2$  $K$-$\pi$ matrix elements,
$\langle \pi \vert O^{\Delta S=1}_{i}\vert K \rangle$. 
A further interesting application is related to hyperon decays. In 
hyperon decays both the parity-odd and parity-even terms contribute. 
For the parity-odd case, as explained in \ref{subsec:renormalization}, there 
are no subtractions induced by the chiral symmetry of the lattice 
action; for the parity-even case, as done before,  we add 
to the relevant correlation function a soft-pion  
(zero four momentum)
field, i.e. a pseudoscalar density summed over $x$, multiplied by 
the factor $2 m$ and replace  the parity-even operator with 
the corresponding  parity-odd one.  Unfortunately,  this method works  only for 
$\Delta I=3/2$ transitions since, in the $\Delta I=1/2$ case, the 
presence of power divergences, due to mixing with lower dimensional 
operators, makes the strategy proposed in this paper almost impossible 
to  implement in practice. 
\section{Conclusion}
With Wilson-like  fermions,  the calculation of the $K^{0}$--$\bar 
K^{0}$ amplitude is complicated by the necessity of accurately  determining 
many mixing coefficients (either perturbatively or 
non-perturbatively), which arise from the explicit chiral 
symmetry breaking in the lattice action. In this paper, we have shown 
that, by using suitable Ward identities,
it  is possible to compute the physical amplitudes without  
any subtraction.  The error  on the final answer is of ${\cal O}(a)$, 
which corresponds to the best accuracy attainable with a perfect 
determination of the mixing coefficients in previous approaches.
The extension of this method to the improved case, i.e. to obtain 
matrix elements with an accuracy of ${\cal O}(a^{2})$,  does not seem 
possible at present, but it is worth being investigated. Our method 
can also be applied to $\Delta I=3/2$ hyperon decays.  
A pioneering numerical calculation of the $K^{0}$--$\bar K^{0}$  
matrix elements using the strategy proposed in this paper is already 
underway.
\section*{Acknowledgements}
G.M.  warmly thank R.~Frezzotti,
S.~Sint and A.~Vladikas for illuminating discussions 
at the 2000 Ringberg workshop.  Their proposal to compute 
the mixing amplitude from the parity-odd operator using twisted-mass 
fermions   combined with a chiral rotation, has triggered the present
investigation. 
We thank C.T.~Sachrajda, S.~Sharpe  and M.~Testa for 
stimulating  discussions and useful comments. 
V. G. has been supported by CICYT under the Grant AEN-96-1718,
by DGESIC under the Grant PB97-1261 and by the Generalitat Valenciana
under the Grant GV98-01-80.  We acknowledge the M.U.R.S.T.
and the INFN for partial support.

\end{document}